\documentclass[aps,twocolumn,prb,showpacs,reprint,amsmath,amssymb,superscriptaddress]{revtex4-1}

\usepackage{graphicx}
\usepackage{dcolumn}
\usepackage{bm}
\usepackage{array}

\begin{document}

\title{The origin of gate hysteresis in p-type Si-doped AlGaAs/GaAs heterostructures}

\author{A.M.~Burke}
\affiliation{School of Physics, University of New South Wales,
Sydney NSW 2052, Australia}

\author{D.~Waddington$^{\dag}$}
\affiliation{School of Physics, University of New South Wales,
Sydney NSW 2052, Australia}

\author{D.~Carrad$^{\dag}$}
\affiliation{School of Physics, University of New South Wales,
Sydney NSW 2052, Australia}

\author{R.~Lyttleton}
\affiliation{School of Physics, University of New South Wales,
Sydney NSW 2052, Australia}

\author{H.H.~Tan}
\affiliation{Department of Electronic Materials Engineering,
Research School of Physics and Engineering, The Australian National
University, Canberra ACT 0200, Australia}

\author{P.J.~Reece}
\affiliation{School of Physics, University of New South Wales,
Sydney NSW 2052, Australia}

\author{O.~Klochan}
\affiliation{School of Physics, University of New South Wales,
Sydney NSW 2052, Australia}

\author{A.R.~Hamilton}
\affiliation{School of Physics, University of New South Wales,
Sydney NSW 2052, Australia}

\author{A.~Rai}
\affiliation{Angewandte Festk\"{o}rperphysik, Ruhr-Universit\"{a}t
Bochum, D-44780 Bochum, Germany}

\author{D.~Reuter}
\affiliation{Angewandte Festk\"{o}rperphysik, Ruhr-Universit\"{a}t
Bochum, D-44780 Bochum, Germany}

\author{A.D.~Wieck}
\affiliation{Angewandte Festk\"{o}rperphysik, Ruhr-Universit\"{a}t
Bochum, D-44780 Bochum, Germany}

\author{A.P.~Micolich}
\email{adam.micolich@nanoelectronics.physics.unsw.edu.au}
\affiliation{School of Physics, University of New South Wales,
Sydney NSW 2052, Australia}

\date{\today}

\begin{abstract}

Gate instability/hysteresis in modulation-doped p-type AlGaAs/GaAs
heterostructures impedes the development of nanoscale hole devices,
which are of interest for topics from quantum computing to novel
spin physics. We present an extended study conducted using
custom-grown, matched modulation-doped n-type and p-type
heterostructures, with/without insulated gates, aimed at
understanding the origin of the hysteresis. We show the hysteresis
is not due to the inherent `leakiness' of gates on p-type
heterostructures, as commonly believed. Instead, hysteresis arises
from a combination of GaAs surface-state trapping and charge
migration in the doping layer. Our results provide insights into the
physics of Si acceptors in AlGaAs/GaAs heterostructures, including
widely-debated acceptor complexes such as Si-X. We propose methods
for mitigating the gate hysteresis, including poisoning the
modulation-doping layer with deep-trapping centers (e.g., by
co-doping with transition metal species), and replacing the Schottky
gates with degenerately-doped semiconductor gates to screen the
conducting channel from GaAs surface-states.
\end{abstract}


\maketitle

\section{introduction}

The modulation-doped AlGaAs/GaAs heterostructure is a materials
platform of great importance to the study of nanoscale electronic
devices with quantum mechanical functionalities, and their
development towards future technologies.~\cite{CapassoPT90,
GoodnickIEEE03} Studies of the two-dimensional electron gas (2DEG)
formed in an AlGaAs/GaAs heterostructure at temperatures below $4$~K
have met with great success, both in terms of the novel physics of
2D electrons,~\cite{AndoRMP82, StormerRMP99} and as an underpinning
technology for quantum wires,~\cite{BeenakkerSSP91, MicolichJPCM11}
quantum dots,~\cite{KouwenhovenRPP01, vanderWielRMP03, HansonRMP07}
and other ballistic transport devices.~\cite{GoodnickIEEE03} Devices
based on two-dimensional hole gases (2DHGs) have received less
attention; this is not from a lack of interesting physics. The
higher effective mass of holes leads to stronger carrier
interactions, making 2DHGs of interest for studies of the
metal-insulator transition~\cite{HaneinPRL98, PapadakisSci99} and
bilayer quantum Hall effect.~\cite{TutucPRL04, ClarkePRB05}
Additionally, the spin-$\frac{3}{2}$ nature of holes, arising from
strong spin-orbit interactions, has driven interest in novel
phenomena in 2DHGs, such as $g$-factor
anisotropy~\cite{WinklerPRL00} and anomalous
spin-polarization,~\cite{WinklerPRB05} as well as studies of the
$0.7$ plateau in quantum point contacts (QPCs),~\cite{DanneauPRL08}
the quantum dot Kondo effect~\cite{KlochanPRL11} and Berry's phase
in Aharonov-Bohm rings.~\cite{YauPRL02} The reduced hyperfine
interaction for holes in GaAs~\cite{KeaneNL11} leads to reduced
spin-decoherence time compared to electrons,~\cite{HeissPRB07,
GerardotNat08} making GaAs hole quantum dots of interest for quantum
computing.~\cite{BulaevPRL05}

Studies of low-dimensional hole devices are impeded by difficulties
in making devices with high electronic stability and low noise/drift
under electrostatic gating. Telegraph noise, instability and gate
hysteresis were particularly problematic in initial attempts to
realize hole QPCs~\cite{ZailerPRB94, DaneshvarPRB97, RokhinsonSM02}
using Si-doped (311)A-oriented 2DHGs. Similar issues were reported
for gated 2DHGs in C-doped (100) AlGaAs/GaAs
heterostructures.~\cite{GrbicAPL05, GrbicPhD07, GerlJCG07} Numerous
explanations have been offered including surface diffusion of ohmic
contact metal producing a low mobility layer at or close to the
surface,~\cite{ZailerPhD94} charge transfer to states either at the
interfaces or in the semiconductor,~\cite{DaneshvarPhD98} carrier
trapping in deep acceptor levels or in insulating parallel doping
layers,~\cite{GerlJCG07} or that metallic gates on p-GaAs are
`inherently leaky' because of a reduced Schottky barrier relative to
n-GaAs.~\cite{GrbicAPL05, GrbicPhD07, CsontosAPL10} Firmly
establishing the origin of the gate instability/hysteresis will
contribute towards the development of improved materials and devices
for the study of low-dimensional hole systems.

Here we show that gate instability/hysteresis in p-type
modulation-doped AlGaAs/GaAs heterostructures is caused by a complex
interplay between surface-state trapping and gate-induced charge
migration within the doping layer. We focus here on Si-doped
(311)A-oriented heterostructures, but similar physics may occur in
C-doped (100)-oriented heterostructures also. We use three different
experimental approaches: first, we rule out direct charge leakage
between the gate and semiconductor by studying devices where an
insulator layer is deposited underneath the gates; two insulators,
Al$_{2}$O$_{3}$ and polyimide, were investigated. We find that
insulating the gates does not eliminate the hysteresis, instead, it
makes the hysteresis significantly worse. A recent study of a QPC
featuring HfO$_{2}$-insulated gates on a C-doped (100) AlGaAs/GaAs
heterostructure~\cite{CsontosAPL10} also shows that hysteresis can
remain despite insulating the gates. In our experiment, the
semiconductor surface/gate interface is the only aspect that is
different between the insulated and uninsulated gate devices; they
are on the same heterostructure with the same doping. The resulting
large difference in hysteresis points to surface-states as an
important contributing factor. Hence our second approach was to
investigate surface effects by studying matched AlGaAs/GaAs
heterostructures, both with and without an Al$_{2}$O$_{3}$ gate
insulator layer, and Schottky-gated p-type heterostructures with
(NH$_{4}$)$_{2}$S$_{x}$ surface passivation
treatment~\cite{SandroffAPL87, YablonovitchAPL87, LebedevPSS02}
performed prior to gate deposition. In contrast to hole devices, the
addition of the insulator does not induce hysteresis for electron
devices but alters the pinch-off voltage. Sulfur passivation does
not bring consistent improvement in p-type devices, despite
increasing the photoluminescence yield equivalently on (100) and
(311) surfaces.~\cite{CarradMIP12} In the one instance where
passivation did bring significant improvement, hysteresis was still
observed. This led us to suspect charge migration in the doping
layer. Hence our third approach was to investigate dopant effects
through variable temperature studies. In particular, we compare the
hysteresis in hole devices to electron devices at elevated
temperatures $T \sim 130$~K, where deep donors known as DX
centers~\cite{MooneyJAP90} begin to detrap,~\cite{ScannellPRB12}
allowing charge to migrate between Si dopant sites. We observe gate
hysteresis in electron devices at $130$~K very similar to that in
hole devices at $T < 4$~K. We also show that the hysteresis can be
reduced in hole devices by reducing the thickness of the dopant
layer from $80$~nm to less than $5$~nm ($\delta$-doped). Overall,
our results point to competition between surface-state and dopant
related processes with different time/energy scales as the cause of
the hysteresis, and have wider implications given recent interest in
surface charge as a source of scattering~\cite{MakAPL10} and dopant
charge migration as a source of noise~\cite{BuizertPRL08} in
AlGaAs/GaAs heterostructures and quantum devices.

The paper is structured as follows. Section~II briefly addresses
materials and methods, with an extended discussion presented in
Appendix A. In Section~III we focus on gate leakage and discuss hole
devices featuring Al$_{2}$O$_{3}$ and polyimide gate insulators.
Sections~IV and V concentrate on surface-states and charge migration
in the dopant layer, respectively. Finally, in Section VI, we draw
conclusions based on the overall results and discuss the broader
implications of the work for the study of low-dimensional devices in
AlGaAs/GaAs heterostructures. Appendices B-D contain additional
supporting data.

\section{Materials and Methods}

Five separate AlGaAs/GaAs heterostructures denoted 1-e, 1-h, 2-e,
2-h and 3-h were made for this study; the first four are `matched'
electron (-e) and hole (-h) heterostructures produces by deposition
onto side-by-side halves of 2" diameter GaAs substrate, one
(100)-oriented and the other (311)-oriented. This relies on the
orientation-dependent amphoteric nature of Si dopants in
AlGaAs.~\cite{WangAPL85} The first four wafers feature an $80$~nm
Si-doped AlGaAs modulation doping layer separated from the 2DHG/2DEG
by $35$~nm of undoped AlGaAs, the fifth features a Si-doping layer
separated from the 2DHG/2DEG by $21$~nm of undoped AlGaAs. These
heterostructures have typical carrier densities $\sim 1.3 \times
10^{11}$~cm$^{-2}$ and mobilities $500,000$~cm$^{2}$/Vs at
temperature $T = 300$~mK. Devices were produced by standard GaAs
device processing methods and involved definition of a Hall bar
structure by wet etching, deposition and annealing of Ohmic contacts
and deposition of Ti/Au gates and interconnects. Al$_{2}$O$_{3}$
insulated gate samples feature a $20$~nm Al$_{2}$O$_{3}$ layer
deposited using atomic layer deposition at $200^{\circ}$C after Hall
bar definition, with access for Ohmic contacts provided using a
buffered HF etch. The polyimide insulated gate sample was produced
using photo-processable polyimide (HD Microsystems) deposited after
the Ohmic contact anneal. Sulfur passivation was performed
immediately before gate deposition by $2$~min immersion in a $0.5\%$
dilution of stock (NH$_{4}$)$_{2}$S$_{x}$ solution prepared by
adding $9.62$~g of elemental sulfur (Aldrich) to $100$~mL of $20\%$
(NH$_{4}$)$_{2}$S solution (Aldrich). The passivation is performed
at $40^{\circ}$C and care is taken to minimize air exposure between
passivation and gate/interconnect deposition. Electrical
measurements at $T \geq 4$~K and $T = 0.25 - 4$~K were obtained
using a liquid helium dip-station and an Oxford Instruments Heliox
$^{3}$He cryostat, respectively. Standard two- and four-terminal ac
lock-in techniques were used to measure the conducting channel's
drain current $I_{d}$, typically with a $100~\mu$V constant voltage
excitation at $73.3$~Hz applied to the source. The gate bias $V_{g}$
was applied using a Keithley 2400 source-measure unit enabling
continuous measurement of gate leakage current $I_{g}$ down to
$100$~pA. More complete details of heterostructure growth, device
fabrication and electrical measurement are given in Appendix A.

\section{Gate leakage and Insulated gate device on p-type heterostructures}

\subsection{Are Schottky-gates on p-type heterostructures really inherently leaky?}

A common explanation for instability and gate hysteresis in
Schottky-gated p-type AlGaAs/GaAs heterostructures is charge leakage
from the metal surface-gate into the heterostructure when the gate
is positively biased to deplete the 2DHG. This often relies on the
argument that Schottky gates on p-type heterostructures are
inherently more leaky than on n-type
heterostructures.~\cite{WilliamsBook90, GrbicAPL05, GrbicPhD07,
CsontosAPL10} Ref.~\cite{GrbicPhD07} suggests this occurs because:
(a) GaAs surface states cause the surface Fermi level to be pinned
slightly closer to the valence band than the conduction band, making
the Schottky barrier lower for p-type
heterostructures,~\cite{WaldropAPL84} and (b) depletion of the hole
gas requires the Schottky gates to be forward biased. The latter
argument is incorrect -- depletion is a reverse bias process whether
the underlying structure is n-type or p-type; the bias convention is
reversed under an inversion in doping type.~\cite{SzeBook02} The
former argument requires caution for two reasons. Firstly, in most
heterostructures the GaAs cap is intrinsic, and the dopants in the
heterostructure, whether n- or p-type, are in the AlGaAs layers more
than $5$~nm beneath the surface. Hence the Schottky barrier is the
same for either doping type, suppressing any relative difference in
gate leakage. Secondly, Schottky barrier measurements are usually
performed at $T = 300$~K, where thermionic emission dominates any
leakage current,~\cite{WaldropAPL84, SzeBook02} whereas most quantum
devices~\cite{CapassoPT90, GoodnickIEEE03, AndoRMP82, StormerRMP99,
BeenakkerSSP91, MicolichJPCM11, KouwenhovenRPP01, vanderWielRMP03,
HansonRMP07, HaneinPRL98, PapadakisSci99, TutucPRL04, ClarkePRB05,
WinklerPRL00, WinklerPRB05, DanneauPRL08, KlochanPRL11, YauPRL02,
HeissPRB07, GerardotNat08, BulaevPRL05, ZailerPRB94, DaneshvarPRB97,
RokhinsonSM02, GrbicAPL05, GrbicPhD07, GerlJCG07, ZailerPhD94,
DaneshvarPhD98, GrbicAPL05, GrbicPhD07, CsontosAPL10, KlochanPRL11,
MakAPL10, BuizertPRL08} are studied at $T = 4$~K, where thermionic
emission is quenched due to its exponential temperature dependence.
This should further suppress the leakage current difference between
Schottky gates on n- and p-type heterostructures.

\begin{figure}
\includegraphics[width=8cm]{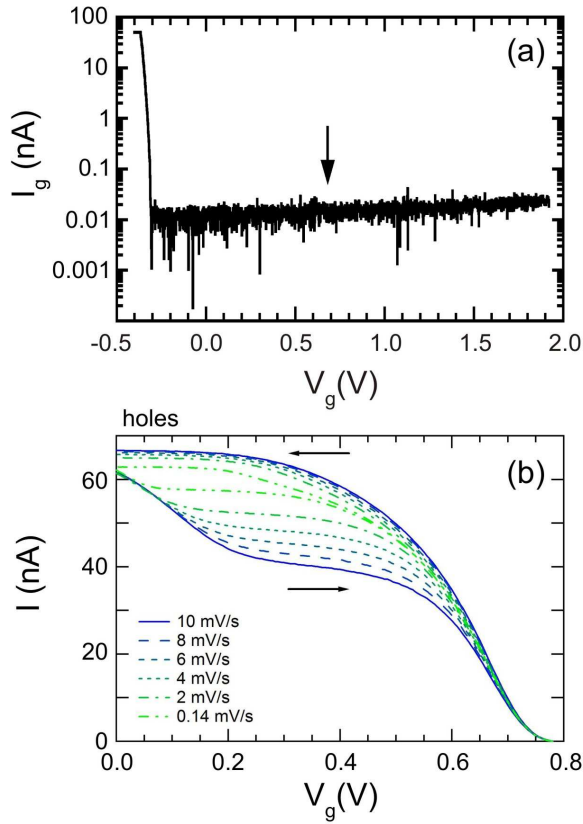}
\caption{(color online) (a) Gate leakage current $I_{g}$ on a
log-axis vs gate bias $V_{g}$ for a Schottky-gated modulation-doped
p-type heterostructure. At positive $V_{g}$, $I_{g}$ remains less
than $50$~pA up to $V_{g} = +2$~V, sufficient to achieve pinch-off
for all uninsulated gate devices studied. In an n-type
heterostructure, gate leakage would normally occur at $V_{g} =
+0.68$~V (indicated by the arrow) and is suppressed for negative
$V_{g}$. (b) Channel current $I_{d}$ vs gate voltage $V_{g}$ at six
different gate sweep rates for Device A. The horizontal arrows
indicate the direction of travel around the hysteresis loop.}
\end{figure}

Thus metal gates on p- and n-type heterostructures should behave
similarly at low temperature aside from a sign-reversal in bias
convention (forward/reverse). To confirm this, Fig.~1(a) shows the
gate leakage current $I_{g}$ versus gate bias $V_{g}$ for a
Schottky-gated p-type heterostructure (This data is presented on a
linear $I_{g}$ axis for comparison in Appendix B). The gate begins
to pass significant current under forward bias conditions at $V_{g}
\approxeq -0.3$~V. Under reverse bias, where 2DHG depletion occurs,
the leakage current gradually rises as $V_{g}$ is made increasingly
positive, with $I_{g} < 50$~pA for the entire range $0 < V_{g} <
+2$~V; this is sufficient to achieve pinch-off in all of our
uninsulated devices. In contrast, a negative $V_{g}$, i.e., forward
bias, produces strong gate leakage at $V_{g} \approxeq -0.3$~V. The
misconception that Schottky gates on p-type AlGaAs/GaAs
heterostructures leak at positive $V_{g}$ may arise from the
knowledge that Schottky gates on n-type AlGaAs/GaAs heterostructures
leak for positive $V_{g} \gtrsim +0.68$~V (indicated by the arrow in
Fig.~1(a)). Reverse bias leakage for n-type heterostructures usually
does not occur until well beyond $V_{g} = -2$~V, consistent with the
reverse bias behavior for the p-type heterostructure in Fig.~1(a).
Despite this, Schottky gates on n-type AlGaAs/GaAs heterostructures
leak relatively tiny amounts of charge under normal
operation.~\cite{CobdenPRL92} This was demonstrated by
Pioro-Ladri\`{e}re {\it et al},~\cite{Pioro-LadrierePRB05} using a
device consisting of a small quantum dot, isolated from the adjacent
2DEG, and coupled to a quantum point contact (QPC) charge sensor
(see Fig.~7 of Ref.~\cite{Pioro-LadrierePRB05}). The QPC detects
charge leaking from the Schottky gates forming the dot, with
extremely small leakage currents $\sim 10^{-20}$~A observed. These
tiny currents are often associated with charge noise in n-type
heterostructures,~\cite{CobdenPRL92, Pioro-LadrierePRB05,
BuizertPRL08} but are not commonly known to produce strong gate
hysteresis in n-type heterostructures to the level seen in p-type
devices; the gates in electron devices are generally very stable
(see Section IV-A). Given this, it is not possible using Fig.~1(a)
and the earlier discussion alone to definitively rule out gate
leakage as a possible cause of gate hysteresis/instability in p-type
heterostructures. The ultimate test, insulating the gates from the
heterostructure, will be presented in Section III-C; first we
characterize the hysteresis we observe in Schottky-gated 2DHGs.

\subsection{Hysteresis in a Schottky-gated p-type heterostructure}

Device A has Ti/Au gates deposited directly on the surface of 1-h.
Figure~1(b) shows the measured drain current $I_{d}$ versus gate
voltage $V_{g}$ from Device A at six $V_{g}$ sweep rates between $0$
and $10$~mV/s. Hysteresis occurs for all $V_{g}$, but is most
prominent for $0.1 < V_{g} < 0.5$~V. Note that $I_{d}$ reflects the
channel conductivity, which can vary due to changes in either
carrier density or mobility. Hall measurements with a small
perpendicular magnetic field applied versus $V_{g}$ show a
qualitatively identical hysteresis to that in Fig.~1(b), confirming
that changes in $I_{d}$ with $V_{g}$ are predominantly density
related. The direction of travel around the hysteresis loop
(indicated by horizontal arrows in Fig.~1(b)) provides important
clues about the origin of the hysteresis. First, it allows the most
obvious cause of apparent hysteresis: recording delay in the
measurement apparatus (sweep lag) to be ruled out. Any delay on the
upsweep to positive $V_{g}$ causes a given depletion to occur at a
higher apparent $V_{g}$, on the downsweep to $V_{g} = 0$,
reaccumulation of carriers occurs at a lower apparent $V_{g}$. The
net result is a clockwise hysteresis loop -- the direction of travel
in Fig.~1(b) is clearly counterclockwise. Another key characteristic
of sweep lag is that the hysteresis is strongest where the
derivative $dI_{d}/dV_{g}$ is greatest. In Fig.~1(b) the hysteresis
is strongest where $dI_{d}/dV_{g}$ is the smallest (see Appendix C).
These two observations confirm that the hysteresis is not an
instrumental issue; it instead originates within the device.

The effect of gate leakage depends on where the charge leaks to. In
an entirely dc measurement, charge leaking directly to the 2DHG
would add/subtract from the channel current. This would modify the
$I_{d}$-$V_{g}$ characteristics but should not produce hysteresis.
Here we measure $I_{d}$ using an ac lock-in technique; any gate
leakage direct to the 2DHG does not appear in the measured $I_{d}$
versus $V_{g}$ unless it is an ac current at the reference
frequency. An alternative is that charge leaks into trap states
between the gate and 2DHG; this will produce hysteresis if the trap
time is not very small. Because a positive $V_{g}$ is applied to the
gates to deplete the 2DHG, any charge leaking from the gate will be
positive, and its closer proximity to the 2DHG will increase the
depletion at a given $V_{g}$ on the downsweep. Similarly, for a 2DEG
the leaked charge will be negative, and it will also increase
depletion; this is what produces the downward steps in the downsweep
arm of the gate characteristics in Fig.~7(a) of
Ref.~\cite{Pioro-LadrierePRB05}. Returning our attention to holes,
on the upsweep, reaccumulation in the 2DHG will be delayed by the
need for the trapped positive charge to dissipate. This produces a
clockwise hysteresis loop for holes, opposite to that in Fig.~1(b)
and consistent with our argument in Section III-C below that gate
leakage does not cause the hysteresis. Note that gate leakage
corresponds to an counterclockwise hysteresis loop for electrons due
to the hysteresis loop mirroring about $V_{g} = 0$ for carrier sign
inversion; this is exactly the loop direction obtained in Fig.~7 of
Ref.~\cite{Pioro-LadrierePRB05} where gate leakage does cause the
hysteresis. The steps in Ref.~\cite{Pioro-LadrierePRB05} arise from
the high sensitivity of their measurement configuration, single/few
trap resolution will not be observed in the large area gate devices
studied here. Note that the hysteresis loop for the hole QPC in
Ref.~\cite{CsontosAPL10} is counterclockwise, opposite the direction
expected for hysteresis due to gate leakage.

We will present a possible explanation for the counterclockwise loop
direction in Section III-D, first, we continue towards ruling out
gate leakage as the cause of the hysteresis by looking at
insulated-gate devices.

\begin{figure}
\includegraphics[width=8cm]{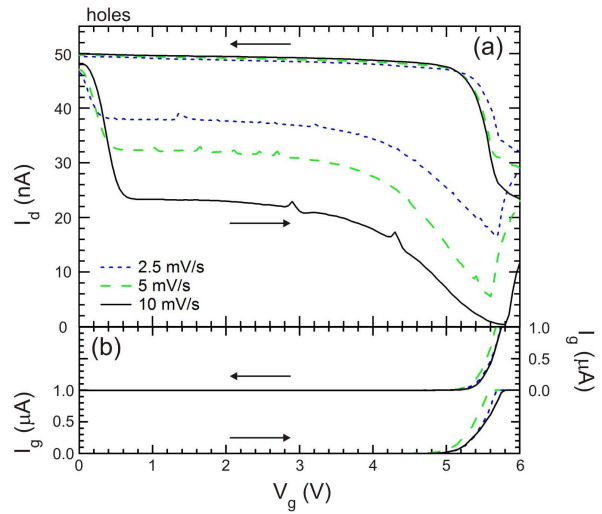}
\caption{(color online): (a) Channel current $I_{d}$ and (b) gate
leakage current $I_{g}$ versus gate voltage $V_{g}$ at three
different gate sweep rates for Device B ($20$~nm Al$_{2}$O$_{3}$
layer on 1-h). Horizontal arrows indicate sweep direction. In (b)
the left/right axis and lower/upper data are for the up/down sweep,
respectively. Once $I_{g}$ reaches $1~\mu$A the gate voltage source
implements current limiting, holding $V_{g}$ fixed. Hence for $V_{g}
\gtrsim +5.6$~V the data in (a) should be considered as $I$ versus
time $t$ at fixed $V_{g}$, with each $0.2$~V minor tick in $V_{g}$
corresponding to $80$, $40$ and $20$~s for sweep rates of $2.5$, $5$
and $10$~mV/s, respectively.}
\end{figure}

\subsection{Hysteresis in an Al$_{2}$O$_{3}$ insulated-gate p-type heterostructure}

Figure~2(a/b) shows $I_{d}$ and $I_{g}$ versus $V_{g}$ for Device B,
with Ti/Au gates insulated by a $20$~nm Al$_{2}$O$_{3}$ layer. We
would expect hysteresis to be heavily suppressed in this device if
it is caused by gate leakage. Instead the hysteresis becomes
stronger, and we need to drive $V_{g}$ to Al$_{2}$O$_{3}$ insulator
breakdown to even approach pinch-off. A higher pinch-off voltage is
expected; Device B has increased gate-2DHG separation and an extra
dielectric layer. We find a conversion factor $V^{Ins}_{g} = 1.21
V_{g}$ relating the insulated and uninsulated gate biases
$V^{Ins}_{g}$ and $V_{g}$ using a parallel-plate capacitor model,
assuming GaAs, AlGaAs and Al$_{2}$O$_{3}$ dielectric constants of
$12.9$, $12.0$ and $9.3$, respectively. It is clear from Figs.~1(b)
and 2(a) that the shift in pinch-off bias from $+0.77$~V to
$+5.8$~V, i.e., by a factor of $7.53$, far exceeds that expected
from simply adding $20$~nm of Al$_{2}$O$_{3}$. To confirm that
Device B's large pinch-off voltage is not a fabrication problem, we
measured a second device differing only in top-gate pattern/area
(see Appendix D). This gave similar behavior, with a large $V_{g}
\sim +5$~V required to achieve even modest depletion ($\sim 22\%$
reduction in $I_{d}$).

The features in Figs.~1(b) and 2(a) are qualitatively identical, but
we highlight two places where the quantitative differences are
substantial. Firstly, the current plateau at intermediate $V_{g}$ is
much longer for Device B, delaying pinch-off accordingly. This
creates the illusion that the initial drop in $I_{d}$ in the upsweep
is steeper; the opposite is true with the initial drop in Fig.~1(b)
complete by $V_{g} = +0.1 - +0.2$~V whereas the same drop in $I_{d}$
takes $V_{g} = +0.4 - +0.6$~V in Fig.~2(a). Secondly, there is a
distinct asymmetry in the sweep-rate dependence of upsweeps and
downsweeps: the upsweep path depends heavily on sweep-rate while the
downsweep path is largely independent of sweep-rate. This is also
evident in Fig.~1(b), but is heavily exacerbated by the
Al$_{2}$O$_{3}$ layer.

The insulator-semiconductor interface can have a radical effect on
performance in devices with a shallow conducting
channel.~\cite{WilkJAP01, GershensonRMP06, SonnetTED10, MakAPL10}
Oxide insulators are particularly troublesome due to high interface
trap densities; improvement is often obtained using polymeric
insulators e.g., polyimide.~\cite{VeresCM04} To explore this we
studied Device C, which features gates insulated with $140$~nm of
polyimide. This gave qualitatively similar results to Device B (see
Appendix D). Pinch-off cannot be achieved in Device C either,
largely due to the much lower breakdown voltage $V_{g} \gtrsim
+2.8$~V for the polyimide layer. At breakdown, $I$ has only fallen
by $\sim 20\%$ of its $V_{g} = 0$ value.

\subsection{Possible explanations for this form of hysteresis}

A possible explanation for an counterclockwise hysteresis loop and
its particular shape in Fig.~1(b) is the gradual
population/depopulation of a layer of {\it net} negative charge
between the gate and 2DHG (n.b., the charge on the gate is positive,
this cannot be gate leakage). On the upsweep, the positive $V_{g}$
places positive charge on the gate, and this ideally results in
depletion of the 2DHG (i.e., reduced carrier density). But, if
adding positive charge to the gate instead results in net negative
charge accumulation between gate and 2DHG, then depletion stalls,
producing a reduced transconductance $|dI_{d}/dV_{g}|$ and perhaps
even an $I_{d}$ plateau. The link between a plateau in $I_{d}$
versus $V_{g}$ and stalling of 2DHG depletion was confirmed by a
corresponding plateau in Hall measurements of the 2DHG. If the
capacity for net negative charge accumulation is finite, depletion
eventually resumes, leading to pinch-off ($I_{d} = 0$). On the
downsweep, positive charge is gradually removed from the gate,
leading to 2DHG repopulation and loss of the accumulated net
negative charge. If the net negative charge is held in deep trapping
sites, its loss may be very slow. Repopulation of the 2DHG will
occur first, producing a rapid rise in $I_{d}$ to a level close to
its initial value at $V_{g} = 0$, followed by a long current plateau
extending to $V_{g} = 0$ as the net negative charge is lost.

An immediately obvious mechanism is charge trapping by surface
states; it is well-known in both oxide-insulated III-V
FETs~\cite{WilkJAP01, SonnetTED10} and organic semiconductor
FETs.~\cite{GershensonRMP06} Another possible mechanism is charge
migration in the modulation-doping layer. Here the use of `net'
charge accumulation is deliberate and important. Silicon donors in
the Al$_{x}$Ga$_{1-x}$As modulation doping layer of an n-type
heterostructure can take two different configurations: a shallow
hydrogenic donor occupying a Ga site, or for $x > 0.2$, a metastable
donor where the Si atom is displaced into an interstitial position
along the $\langle111\rangle$ direction by lattice
distortion.~\cite{BuksSST94} The DX centers act as deep traps; when
the device is cooled below $\sim 120$~K, the DX centers capture free
electrons in the doping layer to become DX$^{-}$, locking some
fraction of the remaining hydrogenic donors in a positive charge
state. This `freezing' dopant layer charge is vital to the stability
and reproducibility of the electronic properties of devices based on
Si-doped n-type AlGaAs/GaAs heterostructures at low
temperatures.~\cite{BervenPRB94, ScannellPRB12, SeePRL12}
Comparatively little is known about the defect physics of Si dopants
in (311)-oriented p-type AlGaAs/GaAs heterostructures. In addition
to acting as a simple substitutional acceptor by occupying an As
site, Si has also been suggested to form an acceptor complex/deep
trap known as Si-X.~\cite{MurrayJAP89} Initial
studies~\cite{MurrayJAP89} suggested that Si-X consisted of a
Si$_{As}$ acceptor adjacent to a Ga site vacancy V$_{Ga}$ (subscript
denotes site occupied). Raman and IR spectroscopy regarding the
existence of Si-X in heavily Si-doped (311)A GaAs layers is
controversial, with data suggesting that it does~\cite{KwokJAP92}
and does not~\cite{AshwinJAP94} exist. Further work, suggested
modified structures for Si-X, first as a
V$_{Ga}$-Si$_{As}$-As$_{Ga}$ complex,~\cite{McQuaidJCG93,
NewmanPRB96} later ruled out in favor of a perturbed
Si$_{Ga}$-V$_{Ga}$ center.~\cite{AshwinJAP97, DomkePRB98} These
studies were all for GaAs; the existence and properties of Si-X-like
complexes in Al$_{0.33}$Ga$_{0.67}$As is unknown. We comment further
on possibilities for Si acceptor complexes and deep traps in Section
VI.

Returning to hysteresis mechanisms, if the dopant layer traps are
shallow or thermal energy is sufficient for a high detrapping rate,
charge can migrate in response to the balance of charge between the
gate and 2DHG. For example, when $V_{g} = 0$ it is energetically
more favorable for the 2DHG side of the doping layer to be net
negative and the gate side net positive. At pinch-off, the
positively charged gate and depleted 2DHG favour the opposite (2DHG
side net positive/gate side net negative). This gate-induced `tidal
flow' of doping layer charge is equivalent to negative charge
accumulation between gate and 2DHG towards generating an
counterclockwise hysteresis loop, as demonstrated in Section V.

\section{The influence of the heterostructure surface}

\subsection{Hysteresis in Schottky and insulated gate devices on matched n-type heterostructures}

To better understand the comparative behavior of Devices A and B, we
prepared Devices D and E on 1-e: Device D has no Al$_{2}$O$_{3}$
layer, Device E has a $20$~nm Al$_{2}$O$_{3}$ layer grown
simultaneously with that in Device B. Figures~3(a/b) present the
electrical characteristics for Devices D and E. In each case the
apparent hysteresis is minimal and the loop direction is consistent
with sweep lag (n.b., reflection of gate characteristics about
$V_{g} = 0$ due to sign-inversion reverses loop direction -- for
electrons sweep lag goes counterclockwise). On its own, the lack of
device-induced hysteresis in Devices D and E does not allow us to
pinpoint the origin of the hysteresis solely to surface-states or
charge migration, as the shift from 1-h to 1-e entails a change in
both the surface orientation and the dopant physics. While the (100)
GaAs surface consists only of double-dangling bonds, the (311)
surface contains equal densities of single- and double-dangling
bonds.~\cite{BoseJAP88}

\begin{figure}
\includegraphics[width=8cm]{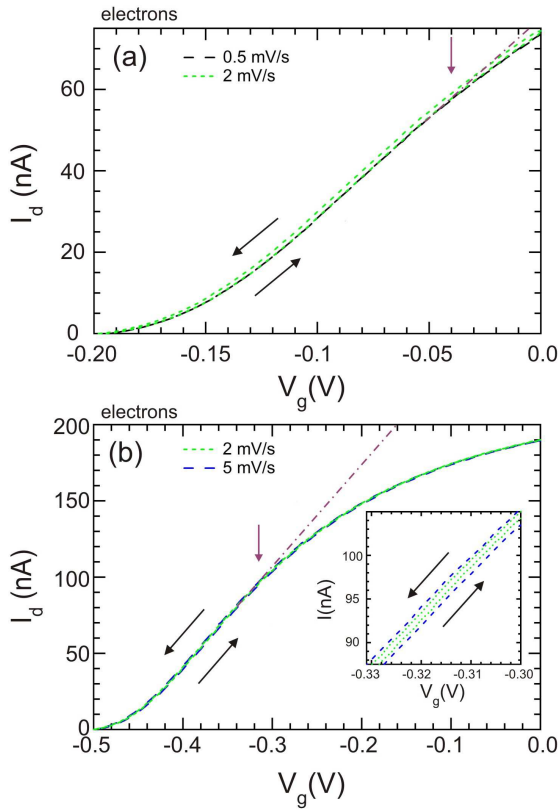}
\caption{(color online): Channel current $I_{d}$ vs gate voltage
$V_{g}$ for devices (a) without (Device D) and (b) with a $20$~nm
Al$_{2}$O$_{3}$ gate insulator (Device E) on 1-e. The inset to (b)
shows a close up of the data in the main panel to highlight the
hysteresis. The arrows indicate hysteresis loop direction. The
purple dot-dashed lines are guides to the eye highlighting the
low-bias non-linearity.}
\end{figure}

Figure~3 has two notable features: the first is pinch-off bias,
which increases from $-0.20$~V to $-0.50$~V upon addition of the
Al$_{2}$O$_{3}$ layer. This $2.5\times$ increase in pinch-off bias
is less than the $7.53\times$ found for holes, but more than double
the $1.21\times$ expected from a parallel-plate capacitor model. The
second feature is the distinct non-linearity in $I_{d}$ versus
$V_{g}$ at low bias in Device E. Although this low bias behavior in
Fig.~3(b) looks different to that in Fig.~2(a), in both cases it
represents a reduced depletion rate for a given change in gate bias,
and may have similar origin. A comparison of Figs.~3(a) and (b)
suggests the low-bias non-linearity arises from the Al$_{2}$O$_{3}$
layer; we believe it is caused by competition between the filling of
surface states and 2DEG depletion. This hypothesis is further
supported by extrapolating the linear trend at moderate $V_{g}$ to
lower $V_{g}$, as per the purple dot-dash lines in Fig.~3(a/b). The
vertical purple arrows indicate where the low-bias non-linearity
ends; $0.16$~V and $0.184$~V to the right of pinch-off in Fig.~3(a)
and (b), respectively. The latter is $1.15\times$ the former, very
close to the $1.21\times$ expected with the Al$_{2}$O$_{3}$ layer.
The $5\%$ difference between the actual value of $0.184$~V for
Device E and the expected value of $0.194$~V is well explained by
the $5\%$ difference in measured $V_{g} = 0$ carrier density between
Devices D and E. The findings above suggest that most 2DEG depletion
occurs in the linear region in Fig.~3(a/b), with surface-state
filling dominant for $V_{g} \gtrsim -0.04$~V in Device D and $V_{g}
\gtrsim -0.32$~V in Device E. Another way to envision this is as
threshold voltage shift induced by the surface-states; as per
oxide-insulated III-V FETs~\cite{WilkJAP01} and organic
semiconductor FETs.~\cite{GershensonRMP06} A final point of note is
that the data in Fig.~3(a) is not linear all the way to $V_{g} = 0$.
This suggests surface-states have a measurable impact on
Schottky-gated devices also. This is not surprising; one naturally
expects a finite surface-state density for uninsulated GaAs surfaces
also.

\subsection{Sulfur passivation of Schottky-gated p-type heterostructures}

One approach to reducing the surface-state density is chemical
passivation, the aim being to remove the native oxide and covalently
satisfy all of the Ga- and As- dangling bonds. This ideally shifts
the surface-states out of the band-gap and into the valence or
conduction bands.~\cite{YablonovitchAPL87, NannichiJJAP88}
Passivation is commonly achieved using aqueous and alcoholic
chalcogenide solutions, particularly those containing sulfur e.g.,
Na$_{2}$S~\cite{SandroffAPL87} or
(NH$_{4}$)$_{2}$S.~\cite{YablonovitchAPL87} A comprehensive review
of chemical passivation of III-V surfaces is provided by
Lebedev.~\cite{LebedevPSS02}

\begin{figure}
\includegraphics[width=8cm]{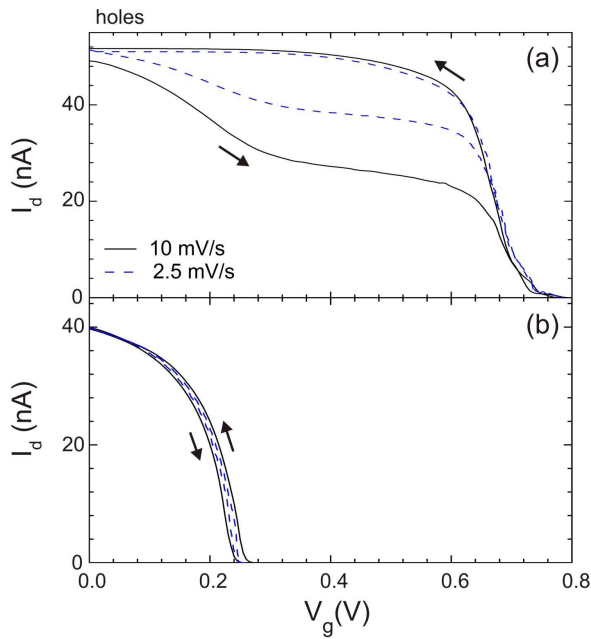}
\caption{(color online): Channel current $I_{d}$ vs gate voltage
$V_{g}$ for (a) Device F on 1-h and (b) Device G on 1-h with sulfur
passivation. The data in (a) is what we typically obtain with sulfur
passivation, even with different treatments.~\cite{CarradMIP12} The
data in (b) is a repeatable measurement, but to date, a
non-reproducible device. We are still working to establish reliable
conditions that produce this outcome. The solid black and dashed
blue traces were obtained at sweep rates of $10$ and $2.5$~mV/s. The
arrows indicate hysteresis loop direction.}
\end{figure}

The best approach to sulfur passivation involves difficult decisions
amongst competing benefits. For example, while Na$_{2}$S treatment
produces a surface passivation that is more robust to light/oxygen
than (NH$_{4}$)$_{2}$S treatment,~\cite{SandroffJVSTB89} the latter
produces surfaces with less O, more S and no traces of
Na.~\cite{BesserJAP89} Alcoholic solutions are more effective than
aqueous solutions,~\cite{BessolovSST98} but alcoholic solutions are
incompatible with photolithography resist, making them difficult it
implement with patterned gates.~\cite{CarradMIP12} The passivation
treatment used here results from an extended study of different
approaches to sulfur passivation of patterned-gate hole devices that
will be reported elsewhere.~\cite{CarradMIP12}

Figure~4(a/b) shows gate hysteresis data from passivated devices on
1-h (Device F/G). We usually obtain data like that in Fig.~4(a) on
including sulfur passivation in the fabrication process; pinch-off
voltages between $+0.57$ and $+0.92$~V are typically obtained
depending on the treatment formulation.~\cite{CarradMIP12} The
Device G data in Fig.~4(b) represents an isolated instance where
elimination of the current plateaus and a much lower pinch-off
voltage $+0.27$~V was observed. Note that a small counterclockwise
hysteresis loop remains, consistent with that obtained when we
examine the hysteresis generated by charge reorganization in the
dopant layer in Section V. It is also consistent with the hysteresis
we observe at $T < 300$~mK and the data in Ref.~\cite{CsontosAPL10}.
Caution is needed with Fig.~4(b) because although the measurement
itself is repeatable, i.e., if we remeasure this device we get the
same result, after several months work we are unable to produce
another device showing the same behavior. The loss of the current
plateaus in Fig.~4(b) bears further discussion. Comparing with
Fig.~4(a) and earlier data, Device G pinch-off occurs before the
current plateau normally begins. The initial depletion is very
strong compared to the other devices; by $V_{g} = +0.25$~V the
current has dropped to zero as opposed to the $30 - 50\%$ found in
other devices without the $V_{g} = 0$ conductivity being
significantly less than normal. The very similar $I_{d}$ at $V_{g} =
0$ for Devices F and G shows that the radically different pinch-off
voltages are not due to a correspondingly large difference in $V_{g}
= 0$ carrier density. Although the rapid initial depletion in Device
G shows a strong reduction in the effect of the surface states on
$I_{d}$ versus $V_{g}$, it is unclear this requires a major change
in the surface state spectrum. A reduction in surface state density
in the tail of the distribution causing the current plateau may
suffice to ensure that depletion is completed before current plateau
onset, giving the radical difference in pinch-off voltage between
Devices F and G.

Putting Device G aside momentarily, our work suggests that
passivation does little to reduce the hysteresis.~\cite{CarradMIP12}
An apparent initial explanation is that the passivation solution is
ineffective on (311) surfaces -- there are no prior studies on this
GaAs surface, development has focussed on the more commonly used
(100), (110) and (111) surfaces.~\cite{LebedevPSS02} We have
performed comparative studies of the efficacy of sulfur passivation
on (100) and (311) GaAs surfaces using photoluminescence
measurements,~\cite{LiuAPL88} and find similar improvement in
photoluminescence yield for both surfaces.~\cite{CarradMIP12} Thus,
while it is evident that passivation significantly affects the
surface states, as expected from earlier work,~\cite{SandroffAPL87,
YablonovitchAPL87, LiuAPL88, NannichiJJAP88, SandroffJVSTB89,
BesserJAP89, BessolovSST98} this does not translate into a
substantial change in the observed hysteresis. A possible
explanation is that the single-dangling bonds present on the (311)
surface interrupt the surface chemistry, reducing passivation
treatment effectiveness. The importance of dangling-bond
presentation to surface chemistry is well known; for example, it
affects the incorporation probability of Si into Al/Ga sites versus
As sites during growth.~\cite{GalbiatiSST96} XPS studies show that
ammonium sulfide treatment of (100) GaAs surfaces leads to disulfide
bridges between adjacent surface As atoms.~\cite{SandroffJVSTB89} It
would be interesting to investigate whether this changes for the
(311) surface; the corresponding effect on surface-state spectrum
could be established using deep level transient spectroscopy
(DLTS).~\cite{LiuAPL88} A focused surface chemistry study may
ultimately reveal a passivation formulation that produces the
improvement found for Device G consistently.

Device G shows that passivation does not eliminate the hysteresis
entirely and that surface-states are not the whole story. There is
additional data supporting this, for example: the addition of the
Al$_{2}$O$_{3}$ layer to a (100)-oriented n-type heterostructure
(Fig.~3) causes a threshold shift consistent with a large change in
surface-state spectrum but does not introduce hysteresis; yet
hysteresis is reported for (100)-oriented p-type
heterostructures~\cite{GrbicPhD07, GerlJCG07, CsontosAPL10} with a
loop direction (counterclockwise) consistent with our observations.
These results suggest that dopants also play a role in the
hysteresis; we now explore this possibility.

\section{Evidence for the role of dopants}

The key to the remarkable stability and performance of n-type
AlGaAs/GaAs heterostructures is the DX center, a deep trap
consisting of a lattice-distorted Si$_{Ga}$ site.~\cite{MooneyJAP90}
If an n-type heterostructure is warmed above $\sim 120$~K, the DX
centers begin releasing their electrons allowing the doping layer
charge distribution to change.~\cite{BuksSST94, BervenPRB94,
ScannellPRB12} This provides an ideal system for studying whether
charge motion in the doping layer generates hysteresis similar to
that observed for p-type heterostructures. Gate hysteresis data from
Device D at $T = 120$ and $130$~K is shown in Figure~5(a/b). The
$120$~K data looks like an intermediate between that in Fig.~3(a)
and Fig.~3(b); the pinch-off voltage is slightly higher ($-0.3$~V
rather than $-0.2$~V) due to the increased temperature. While the
apparent hysteresis in Fig.~3 runs counterclockwise, indicative of
sweep-lag, the hysteresis in Fig.~5(a) runs clockwise, the direction
corresponding to that observed in p-type heterostructures. The
hysteresis becomes more pronounced at $130$~K and its shape is
interesting. As Fig.~5(c) illustrates, if one takes the
characteristic shape obtained for holes, e.g., Fig.~1(b), removes
the current plateaus (green dotted segments), closes the gap and
mirrors about $V_{g} = 0$ to account for carrier charge inversion,
then a hysteresis loop with the same shape as that in Fig.~5(a) is
obtained. The loss of the current plateaus is consistent with the
lack of hysteresis in Fig.~3(a/b). This suggests the current
plateaus are specific to holes, and likely a surface-state effect,
consistent with Figs.~2 and 4(b). A notable feature of Fig.~5(a) is
that the slow cycle (dashed blue trace) has a much higher pinch-off
voltage than the fast cycle (solid black trace); this also occurs
for p-type heterostructures at $T < 1$~K, as discussed below.

\begin{figure}
\includegraphics[width=8cm]{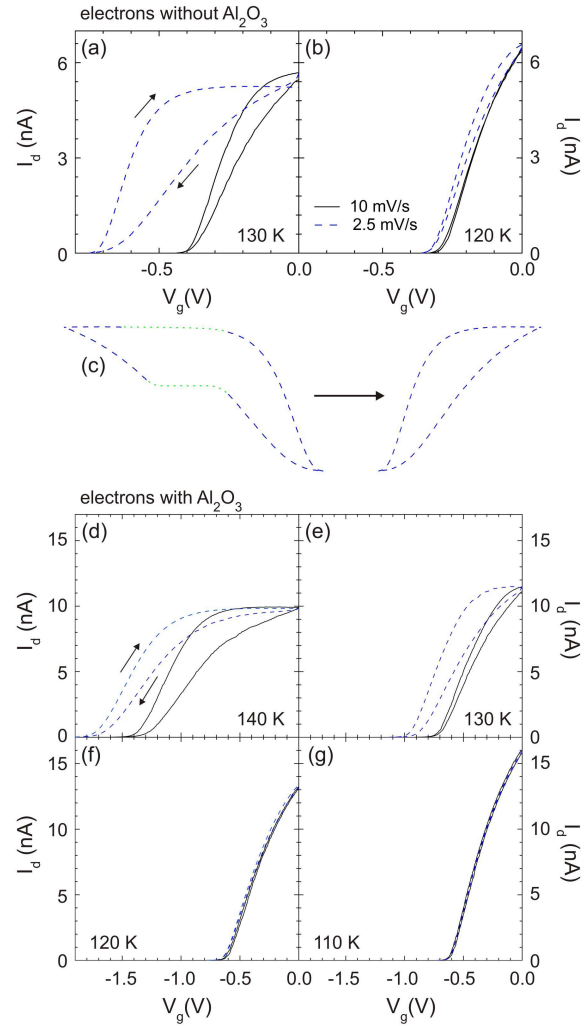}
\caption{(color online): (a,b) Channel current $I_{d}$ vs gate
voltage $V_{g}$ for Device C on 1-e without Al$_{2}$O$_{3}$ at $T =
$ (a) $130$~K and (b) $120$~K. (c) A schematic illustrating an
evident relationship between the hysteresis loop shape in hole
(left) and electron devices (right), as discussed in the text. (d-g)
$I$ vs $V_{g}$ for Device D on 1-e with a $20$~nm Al$_{2}$O$_{3}$
layer at $T = $ (d) $140$~K, (e) $130$~K, (f) $120$~K, and (g)
$110$~K. In (a,b,d-g) the solid black and dashed blue traces were
obtained at sweep rates of $10$ and $2.5$~mV/s. The arrows indicate
hysteresis loop direction. }
\end{figure}

Unfortunately, the gates in Device D begin to leak directly to the
2DEG for $T > 130$~K, preventing higher temperature measurements. To
go higher in $T$ and further explore this behavior, we performed the
same study using Device E (Fig.~5(d-g)). Starting at $T = 110$~K, no
hysteresis appears; the data resembles that in Fig.~3(b) from Device
D, albeit with a slightly higher pinch-off voltage. As $T$ is
increased very similar hysteresis to that in Device D emerges for $T
> 120$~K. The pinch-off voltage increases markedly with $T$, this
limited the measurements to $T \leq 140$~K. Beyond this, the
pinch-off voltage exceeds Al$_{2}$O$_{3}$ layer breakdown causing
gate leakage. Device E gives similar results to Device D, both in
terms of the overall hysteresis loop shape and direction of travel,
and the trend for higher pinch-off voltage at lower sweep-rate.

\begin{figure}
\includegraphics[width=8cm]{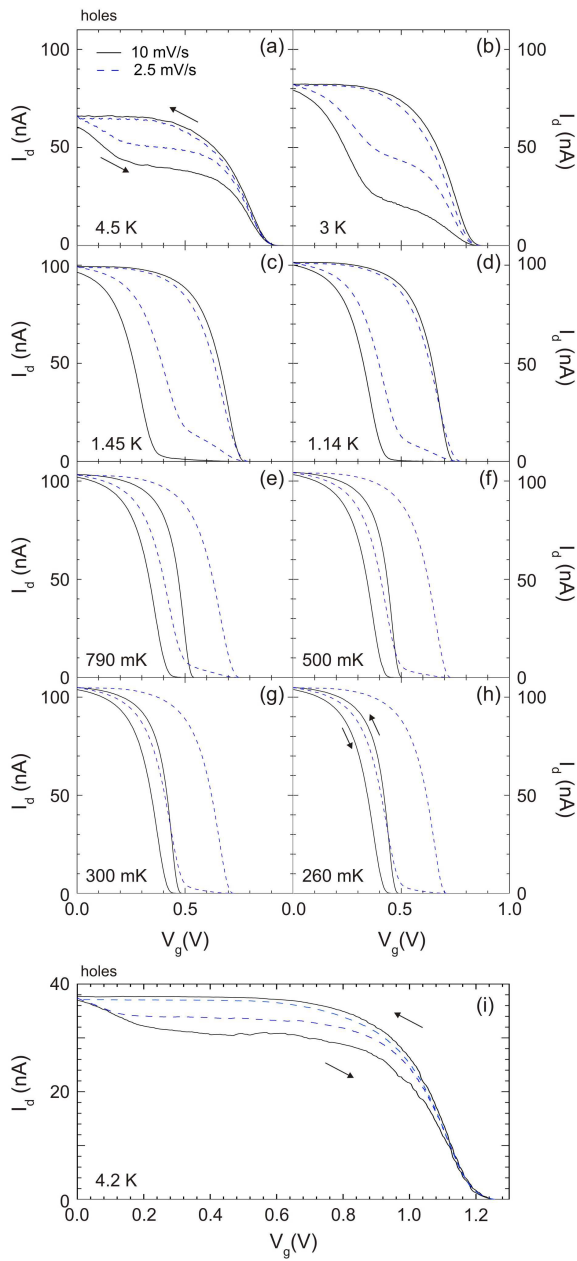}
\caption{(color online): Channel current $I_{d}$ versus gate voltage
$V_{g}$ for Device F on 2-h without Al$_{2}$O$_{3}$ at $T =$ (a)
$4.5$~K and (b) $3.0$~K, (c) $1.45$~K, (d) $1.14$~K, (e) $790$~mK,
(f) $500$~mK, (g) $300$~mK and (h) $260$~K. (i) $I_{d}$ versus
$V_{g}$ for Device H on the $\delta$-doped heterostructure 3-h
without Al$_{2}$O$_{3}$/passivation at $T = 4$~K. The solid black
and dashed blue traces were obtained at sweep rates of $10$ and
$2.5$~mV/s. The arrows indicate hysteresis loop direction.}
\end{figure}

In Fig.~6(a-h) we show the corresponding behavior for holes. As $T$
is reduced from $T = 4.5$~K to $T = 260$~mK, two obvious changes
result: first, the current plateau shortens and drops to lower
$I_{d}$. The plateau shortening and lowering are linked -- if
plateau onset shifts to higher $V_{g}$ then the low $V_{g}$
depletion proceeds further before the plateau appears. Second, for
$T < 1.14$~K the pinch-off voltage for the slow trace significantly
exceeds that for the fast trace. A careful inspection of
Figs.~6(c-e) reveals that this caused by loss of the fast trace
current plateau, while the slow trace current plateau remains until
$T < 260$~mK. For the fast trace hysteresis loop at $T < 1.14$~K,
the hysteresis remains despite the loss of the current plateau and
strongly resembles that in Fig.~5 and 4(b). This behavior clearly
indicates that two processes are involved in generating the
hysteresis -- surface-state trapping and charge redistribution
amongst the dopants -- each with different time and energy scales.
Attributing the current plateau to surface state trapping, it
appears that the surface-states `freeze out' at $T < 1$~K leaving
the dopant effects behind. This might explain the lack of hysteresis
in the HfO$_{2}$-insulated hole QPC studied by Csontos {\it et
al}~\cite{CsontosAPL10} compared to our Al$_{2}$O$_{3}$ device (Fig.
2). The data shown in Fig.~3(a) of Ref.~\cite{CsontosAPL10} was
obtained at $T = 100$~mK; here the surface-state component of the
hysteresis is likely quenched leaving only the hysteresis due to
charge migration in the doping layer. This is why the hysteresis
occurs primarily closer to pinch-off, consistent with Fig.~4(b)
where surface-state hysteresis also appears to be quenched. The
hysteresis loop direction in Ref.~\cite{CsontosAPL10} is consistent
with this explanation; data at higher $T$ for this device would be
enlightening. To further demonstrate the influence of dopants in the
hysteresis we discuss one final device. In Figure~6(i) we present
hysteresis data from Device H, made on the $\delta$-doped
heterostructure 3-h with Schottky-gates on an unpassivated surface.
The key difference is that the dopants only have a vertical spread
$< 5$~nm in Device H, compared to $\sim 80$~nm for Devices A-G. The
2DHG is closer to the surface ($47$~nm for 3-h versus $120$~nm for
1-h) and the $5$~nm GaAs cap in 3-h is Si doped unlike
heterostructures 1 and 2. Doping of the cap is normally performed to
improve ohmic contact formation; however, the doping level is
necessarily limited to prevent shorting of gates and Ohmic contacts.
Comparing Fig.~6(i) with Fig.~5(a/b) there are two notable
differences. First and foremost, the hysteresis loop's vertical
extent is reduced from $42\%$ to $17\%$. This reduction is likely
due to the $> 16$-fold reduction in dopant layer spread for Device
H. Second, the pinch-off voltage and width of the current plateau
are increased. A cap doping of $\sim 1-2 \times 10^{18}$~cm$^{-3}$
corresponds to replacing only roughly 1 in every $220,000$ surface
atoms with Si. Considered alongside the sulfur passivation results,
the cap doping is unlikely to be the dominant cause for the increase
in current plateau width. It is more likely that the shallower 2DHG
in 3-h exacerbates the surface-state contribution to the hysteresis.
This would be consistent with scattering studies in shallow undoped
n-type heterostructures where the surface-charge scattering
contribution increased with reduced 2DEG depth.~\cite{MakAPL10}

\section{Conclusions}

We set out to identify the exact origin of the gate hysteresis in
p-type AlGaAs/GaAs heterostructures -- several divergent
explanations exist in the literature and a better understanding will
enable development of low-dimensional hole devices with improved
stability/performance. A commonly accepted explanation is that gates
on p-type heterostructures are inherently leaky. We show that this
not the case; Schottky gates on p-type heterostructures are not
significantly more leaky than Schottky gates on n-type
heterostructures under reverse bias conditions. Hysteresis due to
gate leakage in p-type heterostructures should give a clockwise
hysteresis loop and the hysteresis we observe runs counterclockwise.
We also find that the hysteresis becomes drastically worse rather
than much better if the gates are insulated. We note that hysteresis
was also still observed in QPCs with HfO$_{2}$-insulated gates on
C-doped p-type heterostructures,~\cite{CsontosAPL10} despite these
producing improved tunability compared to equivalent Schottky-gated
devices. The direction of the hysteresis loop in
Ref.~\cite{CsontosAPL10} is counterclockwise and is inconsistent
with gate leakage. The measurements by Csontos {\it et al} were
obtained at very low temperature $T \sim 100$~mK and based on our
data, we propose that this hysteresis may be due to charge
redistribution in the dopant layer.

Our work focussed on the investigation of devices with
insulated/uninsulated gates on custom-grown, matched electron and
hole heterostructures. This relies on Si being an n-type dopant on
(100) substrates and a p-type dopant on (311)A
substrates.~\cite{WangAPL85} Despite the strong hysteresis in p-type
devices at $T = 4$~K, we observe no hysteresis in n-type devices
until the temperature exceeds $T = 120$~K, where the Si DX centers
in the n-type heterostructure begin to detrap and migration of
charge within the doping layer occurs.~\cite{BuksSST94, BervenPRB94,
ScannellPRB12, SeePRL12} The hysteresis that emerges at $T > 120$~K
for electrons bears a strong resemblance to that in holes at lower
temperatures, particularly at $T < 1$~K, where the current plateau
at intermediate $V_{g}$ drops to $I_{d} \sim 0$ and is quenched.
This correspondence, and the lack of hysteresis for electrons at $T
\sim 4$~K, strongly suggests that surface states are not the sole
cause of the hysteresis; migration of charge in the dopant layer is
likely involved as well. We return to surface states following a
discussion of dopants.

Comparatively little is known about the physics of Si acceptors in
(311)A heterostructures. Although Si clearly acts as a
substitutional acceptor, the presence and properties of acceptor
complexes such as Si-X is debated.~\cite{MurrayJAP89, KwokJAP92,
AshwinJAP94, McQuaidJCG93, NewmanPRB96, AshwinJAP97, DomkePRB98} Our
data cannot provide insight at the atomic level, but there are
clearly no acceptor sites in a p-type heterostructure's doping layer
that act like the deep-trapping DX centers in n-type
heterostructures. Regarding Si-X specifically, we can draw two
conclusions: If Si-X exists in Al$_{0.33}$Ga$_{0.67}$As in (311)A
heterostructures, then it must be a very shallow trap (more than 100
times shallower than DX) if present at high density, and only a
deep-trap if it is present at such low density that it cannot
`freeze' the doping layer's charge configuration. While the lack of
DX-like deep traps answers the obvious question of why p-type
heterostructures are so unstable/hysteretic under gating, it is
interesting to invert the thinking and consider instead why n-type
heterostructures are often so impressively stable. The answer is
clear -- the deep-trapping DX centers `lock down' the vast majority
of free charge in the dopant layer -- but inspirational: One way to
stabilize modulation-doped p-type heterostructures may be to
deliberately poison the modulation doping layer with deep trapping
sites, perhaps by engineering the growth conditions to obtain a high
density of deep-trapping Si acceptor complexes, or failing that, by
co-doping with transition metal impurities, e.g., Cu, Fe, Ni or
Zn.~\cite{AuretAPL86, ReemtsmaJAP89, RadueJAP98} Cu is probably more
optimal than Fe or Ni, which may bring magnetic side-effects, and
Zn, which is a rapid diffuser in GaAs and may be incompatible with
Ohmic contact formation.

The changes in hysteresis obtained by changing insulator composition
or surface passivation are small compared to the differences arising
between the presence/absence of a gate insulator. One possible
explanation is that the metal/GaAs interface quenches the
surface-state density, either chemically by forming Ti-O-Ga or
Ti-O-As bonds (or Ti-S-Ga or Ti-S-As bonds), or physically by
providing a nearby high electron density that partially screens the
surface states.~\cite{HoPRB08, HoPRB09} This is in the same spirit
as the addition of a doping layer to compensate the surface-states
in ultra-high mobility n-type AlGaAs/GaAs
heterostructures.~\cite{UmanskyJCG09, RosslerNJP10} This idea could
be applied here with the lightly-doped cap in 3-h replaced by adding
a uniformly doped layer between the cap and modulation-doping for
surface-state compensation.

Another alternative is to abandon Schottky-gates and use a
degenerately-doped cap as a semiconductor gate.~\cite{SolomonEDL84,
KaneAPL98, ClarkeJAP06, KlochanAPL06} This effectively places the
gate underneath the semiconductor surface, enabling the gate to
screen the 2DHG from the surface-states. Indeed, this explains the
high stability and lack of hysteresis in undoped p-type
semiconductor-insulator-semiconductor field-effect transistor
(SISFET) devices.~\cite{KlochanAPL06, KlochanNJP09, KlochanAPL10,
ChenNJP10, KlochanPRL11} There both hysteresis contributions are
dealt with -- the modulation doping is removed and the gate screens
the surface-states. Given the success of undoped SISFETs, one might
ask: Why bother making semiconductor-gated modulation-doped devices?
In undoped SISFETs it is essential that the gate overlaps the Ohmic
contacts; this makes fabrication more difficult and lowers
yield.~\cite{ClarkeJAP06} This overlap is unnecessary in
modulation-doped devices. Hence, if the doping layer charge
migration issue described earlier can be successfully overcome,
modulation-doped semiconductor-gated structures may provide a
formidable platform for studying low-dimensional hole devices.

\begin{acknowledgements}
This work was funded by Australian Research
Council Grants DP0772946, DP0877208, FT0990285 and FT110100072. DR
and ADW acknowledge support from DFT SPP1285 and BMBF QuaHL-Rep
01BQ1035. This work was performed in part using the NSW and ACT
nodes of the Australian National Fabrication Facility (ANFF). We
thank L. Eaves for helpful discussions on DX centers, and L.H. Ho
for the initial measurements that motivated our interest in this
problem. $^{\dag}$: DW and DC contributed equally to the work.
\end{acknowledgements}

\appendix
\section{Detailed Materials and Methods}

Experiments were performed on five separate AlGaAs/GaAs
heterostructures grown by molecular beam epitaxy (MBE), denoted 1-e,
1-h, 2-e, 2-h and 3-h. The first four were produced in two separate
growth runs, each performed onto side-by-side halves of 2" diameter
semi-insulating GaAs substrate, one (100)-oriented and the other
(311)-oriented, to give `matched' electron (-e) and hole (-h)
heterostructures, respectively. Production of electron and hole
heterostructures in a single growth is enabled by the facet-specific
amphoteric nature of Si dopants in AlGaAs (n-type on (100) and
p-type on (311))~\cite{WangAPL85}. Both growth runs have a nominally
identical epilayer structure, the active region consisting of
$650$~nm undoped GaAs, $35$~nm undoped Al$_{0.34}$Ga$_{0.66}$As,
$80$~nm Si-doped Al$_{0.34}$Ga$_{0.66}$As and a $5$~nm undoped GaAs
cap. The $\delta$-doped wafer 3-h begins with $689$~nm undoped GaAs
and $21$~nm undoped Al$_{0.33}$Ga$_{0.67}$As grown at
$690^{\circ}$C. Growth is then interrupted and the substrate cooled
to $580^{\circ}$C. The Si source is opened for $180$~s and then
$5$~nm of undoped Al$_{0.33}$Ga$_{0.67}$As is grown. Growth is
interrupted again to return the substrate to $690^{\circ}$C before
finishing the device with $16$~nm of undoped
Al$_{0.33}$Ga$_{0.67}$As and a $5$~nm GaAs cap. The devices have
typical carrier densities $\sim 1.3 \times 10^{11}$~cm$^{-2}$ and
mobilities $500,000$~cm$^{2}$/Vs at temperature $T = 4$~K; see Table
I for specific values for each heterostructure at $T = 300$~mK using
four-terminal Shubnikov-de Haas and Hall resistivity measurements.
All measurements were performed in the dark. Details for each of the
eight devices studied are presented in Table II.

\begin{table*} \caption{\label{Table1} The Bochum wafer number, electron/hole density and
mobility for the five AlGaAs/GaAs heterostructures studied at $T
\sim 300$~mK.}
\begin{ruledtabular}
\begin{tabular}{ccccc}
Heterostructure & Bochum Wafer Number & Carrier & Density
(cm$^{-2})$ & Mobility
(cm$^{2}$/Vs)\\
\hline
1-e & 13473-e & electrons & $1.20 \times 10^{11}$ & $640,000$\\
1-h & 13473-h & holes & $1.94 \times 10^{11}$ & $805,800$\\
2-e & 13516-e & electrons & $1.46 \times 10^{11}$ & $345,000$\\
2-h & 13516-h & holes & $1.64 \times 10^{11}$ & $1,030,000$\\
3-h & 13483-h & holes & $1.74 \times 10^{11}$ & $1,250,000$\\
\end{tabular}
\end{ruledtabular}
\end{table*}

\begin{table} \caption{Details for the $8$ devices studied. Mod = Modulation doped, $\delta$ = delta-doped}
\begin{ruledtabular}
\begin{tabular}{ccccc}
Device & Heterostructure & Doping & Insulator & Passivation\\
\hline
A & 1-h & Mod & No & No \\
B & 1-h & Mod & Al$_{2}$O$_{3}$ & No \\
C & 2-h & Mod & Polyimide & No \\
D & 1-e & Mod & No & No \\
E & 1-e & Mod & Al$_{2}$O$_{3}$ & No \\
F & 1-h & Mod & No & Yes \\
G & 1-h & Mod & No & Yes \\
H & 3-h & $\delta$ & No & No \\
\end{tabular}
\end{ruledtabular}
\end{table}

Hall bars with a $140$~nm high mesa were produced using
photolithography and a $2:1:20$ buffered HF:H$_{2}$O$_{2}$:H$_{2}$O
wet etch. The buffered HF solution is $7:1$~NH$_{4}$F:HF. The
photolithographically defined Ohmic contacts consist of $150$~nm
AuBe alloy for p-type contacts and a stack containing $5$~nm Ni,
$35$~nm Ge, $72$~nm Au, $18$~nm Ni and $50$~nm Au for n-type
contacts. Contacts were annealed at $490^{\circ}$C for $90$~s
(p-type) and $430^{\circ}$C for $30$~s (n-type). The
photolithographically defined gates consist of $20$~nm Ti and
$100$~nm Au deposited after the ohmic contact anneal. Figure~7(a)
shows an optical micrograph of a completed device.

\begin{figure}
\includegraphics[width=6cm]{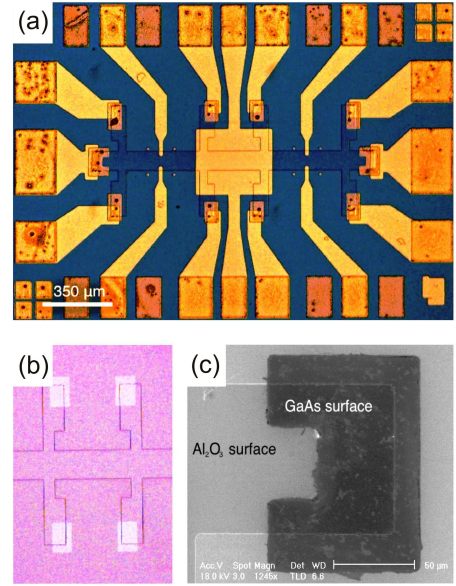}
\caption{(color online): (a) Optical micrograph of a completed
device with $10$ ohmic contacts (three at each end and four near the
middle) and a single top-gate covering the midsection of the Hall
bar with interconnects at top and bottom. The remaining four metal
gates were not used. (b) Nomarski phase-contrast optical micrograph
showing four etched penetrations (off-white) in the $20$~nm
Al$_{2}$O$_{3}$ layer (pink) obtained using a buffered HF etch.
These correspond to the four central ohmic contacts in (a). (c)
Scanning electron micrograph of the etch penetration (dark) in the
Al$_{2}$O$_{3}$ layer (light) for the ohmic contact at the far right
of the Hall bar in (a). (b) and (c) were obtained after the buffered
HF etch and before ohmic metallization.}
\end{figure}

Al$_{2}$O$_{3}$ insulated gate samples were produced by adding the
following steps between mesa etch and ohmic contact metallization. A
$20$~nm Al$_{2}$O$_{3}$ layer was deposited using (CH$_{3}$)$_{3}$Al
and H$_{2}$O gaseous precursors at $200^{\circ}$C in a Cambridge
Nanotech Savannah $100$ Atomic Layer Deposition (ALD) system. Access
to the heterostructure for the ohmic contacts was obtained by a
$60$~s buffered HF etch ($7:1$~NH$_{4}$F:HF in H$_{2}$O) following
photolithographic definition of the contact regions and prior to
metallization. In the absence of H$_{2}$O$_{2}$ this etch
self-terminates, stripping the native GaAs surface oxide in the
process. The openings in the Al$_{2}$O$_{3}$ layer are visible
optically (Fig.~7(b)), but are clearer using scanning electron
microscopy (Fig.~7(c)). Although traces of residual oxide remain,
they do not adversely affect ohmic contact formation.

The polyimide insulated gate sample was produced with an added step
between the ohmic contact anneal and gate deposition. A $140$~nm
layer of patterned polyimide is obtained by spin-coating a diluted
mixture (1:1.4) of photo-processable polyimide (HD Microsystems
HD-4104) in thinner (HD Microsystems T-9039), performing a
$65^{\circ}$C soft-bake for $90$~s, exposing/developing aligned
ohmic contact openings, and finishing with a $250^{\circ}$C
hard-bake for $60$~min under $1$~atm N$_{2}$.

The sulfur passivation treatment was performed immediately before
gate deposition. A stock solution of (NH$_{4}$)$_{2}$S$_{x}$ was
prepared by adding $9.62$~g of elemental sulfur (Aldrich) to
$100$~mL of $20\%$ (NH$_{4}$)$_{2}$S solution (Aldrich) and mixing
until completely dissolved. The passivation solution was a $0.5\%$
dilution of the stock solution in deionized water (Millipore).
Passivation was performed by immersion for $2$~min in $3-5$~mL of
passivation solution heated to $40^{\circ}$C in a water bath,
followed by a deionized water rinse. The sample is stored under
deionized water during transfer to the vacuum evaporator. The total
time between passivation and sample at vacuum was $< 30$~min, with
the sample exposed to air for no longer than a few minutes during
evaporator loading and pump-down.

Electrical measurements at $T \geq 4$~K were obtained using a liquid
helium dip-station, with $T > 4$~K achieved using the natural
stratification of the He atmosphere inside the dewar. Data at $0.25
- 4$~K was obtained using an Oxford Instruments Heliox $^{3}$He
cryostat. Standard two- and four-terminal ac lock-in techniques were
used to measure the conducting channel's drain current $I_{d}$,
typically with a $100~\mu$V constant voltage excitation at $73.3$~Hz
applied to the source. Pinch-off (i.e., $I_{d} = 0$) is interpreted
as $I_{d} < 10$~pA; this triggers the software to commence the
downsweep to avoid pointlessly driving the device beyond pinch-off.
The gate bias $V_{g}$ was applied using a Keithley 2400
source-measure unit enabling continuous measurement of gate leakage
current $I_{g}$ down to $100$~pA. This instrument has a built-in
current limiter, with $V_{g}$ curtailed to keep $I_{g}$ at a
specified limit even if a higher $V_{g}$ if requested.

\section{Leakage characteristics of Schottky gates on p-type heterostructures}

\begin{figure}
\includegraphics[width=8cm]{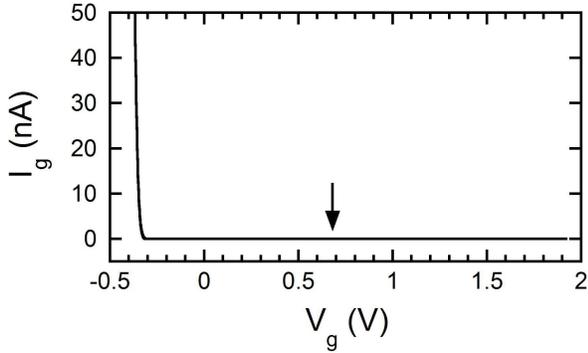}
\caption{(color online): Gate leakage current $I_{g}$ on a linear
axis vs gate bias $V_{g}$ for a Schottky-gated modulation-doped
p-type heterostructure. The data matches that shown in Fig.~1(a). At
positive $V_{g}$, $I_{g}$ remains less than $50$~pA to $V_{g} =
+2$~V, sufficient to achieve pinch-off for all uninsulated gate
devices studied. In an n-type heterostructure, gate leakage would
normally occur at $V_{g} = +0.68$~V (indicated by the arrow) and is
suppressed for negative $V_{g}$.}
\end{figure}

Figure~8 shows the data in Fig.~1(a) plotted on a linear-linear
scale for comparison. The gate leakage current $I_{g}$ is $< 50$~pA
for the entire range $0 < V_{g} < +2$~V.

\section{Analysis of the hysteresis in Fig.~1(b)}

\begin{figure}
\includegraphics[width=8cm]{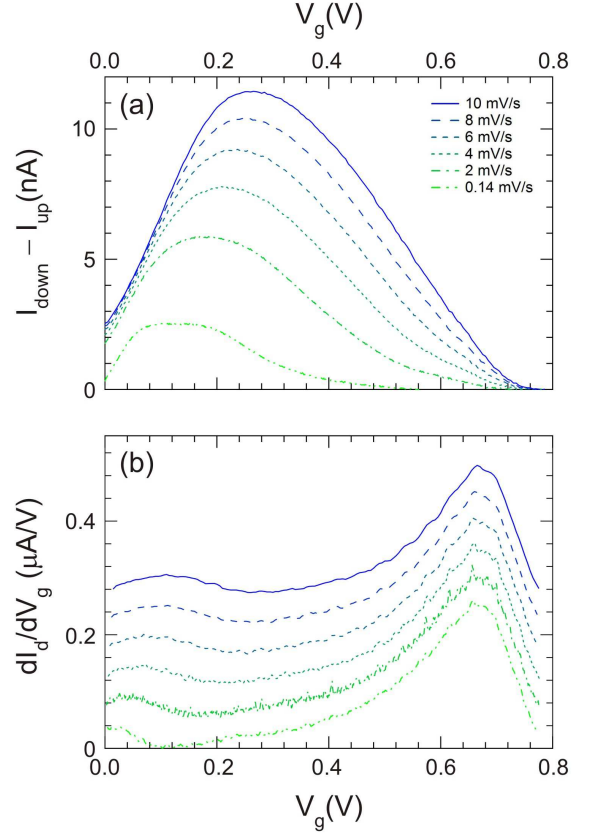}
\caption{(color online): (a) The hysteresis loop vertical extent
$I_{down} - I_{up}$ and (b) the numerical transconductance
$dI/dV_{g}$ for the upsweeps vs gate voltage $V_{g}$. The traces in
(b) are sequentially offset vertically by $+0.05$ with increasing
sweep rate for clarity.}
\end{figure}

Figure~9 shows the vertical extent of the hysteresis loop and the
slope of the upsweeps versus $V_{g}$. The slope curves for the down
sweeps are very similar to those in Fig.~9 except there is less
structure at low $V_{g}$ and the peak at higher $V_{g}$ is sharper
and shifted to more positive $V_{g}$. If the hysteresis is caused by
sweep lag, one would expect the maximum vertical extent in the
hysteresis to coincide in $V_{g}$ with the maximum slope
$dI/dV_{g}$. This is clearly not the case, supporting our assessment
that the hysteresis is not caused by sweep lag.

\section{Reproducibility of the hysteresis result from Device B -- Polyimide insulated gate on a p-type heterostructure}

\begin{figure}
\includegraphics[width=8cm]{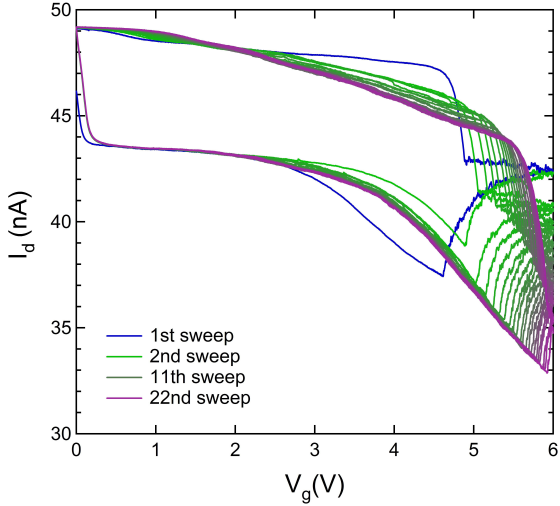}
\caption{(color online): Channel current $I$ vs gate voltage $V_{g}$
for 22 hysteresis loop sweeps of Device B2. The first sweep is show
in blue, the $2$nd to $22$nd sweeps evolve continuously in color
from green to magenta. The sweep rate in each case is $5$~mV/s. The
points of discontinuity at high $V_{g}$ are where gate voltage
current limiting comes on/off for the up/downsweep.}
\end{figure}

Figure~10 shows hysteresis loop data for Device B2, which is
nominally identical to Device B aside from a change in the gate
design used. The gate in Device B2 covers less area. In both cases
the gate does not extend to the edges of the Al$_{2}$O$_{3}$ layer;
hence in neither case is leakage by proximal direct shorting between
gate and ohmic contact. The characteristics are very similar to
those in Device B, and the breakdown voltage is of a similar
magnitude $\sim 5$~V. Repeated sweeps appear to increase the
breakdown voltage, enabling the device to slowly progress towards
pinch-off. The exact cause for this is unclear.

\begin{figure}
\includegraphics[width=8cm]{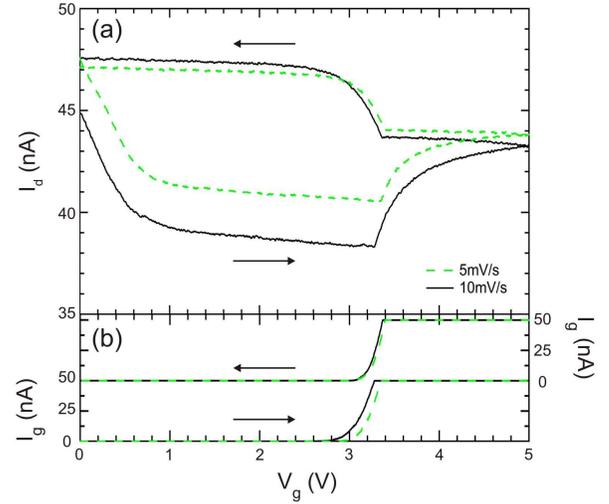}
\caption{(color online):(a) Channel current $I$ and (b) gate leakage
current $I_{g}$ vs gate voltage $V_{g}$ at two different gate sweep
rates for a Device C featuring a $140$~nm polyimide layer on 2-h.
The horizontal arrows indicate the direction of travel around the
hysteresis loop. In (b) the left/right axis and lower/upper set of
data are for the sweep to/from positive $V_{g}$, respectively. Note
that once $I_{g}$ reaches $50$~nA the gate voltage source implements
current limiting by holding $V_{g}$ fixed. Hence for $V_{g} \gtrsim
+3.2$~V the data in (a) should be considered as $I$ versus time $t$
at fixed $V_{g}$. To aid in converting the data at $V_{g} \gtrsim
+3.2$~V into time, each $0.2$~V minor tick in the figure corresponds
to $40$ and $20$~s for sweep rates of $5$ and $10$~mV/s,
respectively.}
\end{figure}

The radical effect that the composition and properties of the
insulator-semiconductor interface can have on transistor performance
is well-known for organic FETs.~\cite{VeresCM04} Due to high
interface-state densities, inorganic oxide insulators are
particularly troublesome, causing significant hysteresis and shifts
in threshold voltage. Polymeric insulators such as polyimide often
bring significant improvement. To explore whether the current
plateau and difficulty in attaining pinch-off in Device B is tied to
insulator composition we studied Device C containing gates insulated
with a $140$~nm thick polyimide layer on 2-h. Figures~11(a/b) show
$I$ and $I_{g}$ versus $V_{g}$ for this device. Pinch-off cannot be
achieved in Device C, largely due to the much lower breakdown field
for the polyimide insulator. Breakdown occurs for $V_{g} \gtrsim
+2.8$~V, and once $I_{g}$ reaches $50$~nA at $V_{g} \sim +3.4$~V the
voltage source holds the gate bias fixed, as in Fig.~2. Thus the
data at $V_{g} \gtrsim 3.4$~V in Fig.~11(a) should instead be
considered as $I$ versus time $t$ at constant $V_{g}$. Here each
minor sub-tick corresponds to a time of $10$~s for the $10$~mV/s
trace and $20$~s for the $5$~mV/s trace. We show the data obtained
beyond pinch-off in Fig.~11 to better facilitate comparison with
Fig.~2, and to highlight the equilibration behavior that occurs when
a sweep is stopped. The $I$ versus $V_{g}$ characteristics for
Device C bear a striking resemblance to those of Device B (Figs.~2
and 10). On the upsweep there is an initial drop in current that
plateaus for intermediate $V_{g}$, followed by a recovery in $I$
when $V_{g}$ is held constant. This is consistent with slow
accumulation of net negative charge between the gate and 2DHG, as
per the explanation for hysteresis loop shape/direction in Section
III-D. The downsweep gives an initial rapid rise in $I$ that
plateaus as $V_{g}$ approaches zero again, and as in Fig.~2, there
is a distinct asymmetry between the upsweeps and downsweeps
regarding the dependence of the path taken on sweep rate. Pinch-off
occurs beyond $V_{g} \sim +3.4$~V, thus the shift in pinch-off bias
must exceed $\sim 3.4/0.77 = 4.41$. The capacitive conversion factor
for this device $V^{Ins}_{g} = 5.18V_{g}$ assuming $140$~nm of
polyimide with dielectric constant $3.36$, giving an anticipated
pinch-off voltage $\sim +4.15$~V.

\end{document}